\documentclass{WileyMSP-template}
\usepackage{dcolumn}
\usepackage{bm}
\usepackage{braket}
\usepackage{color}
\usepackage{accents}
\usepackage{mathrsfs}
\usepackage{bm}
\usepackage{here}
\usepackage{longtable}
\usepackage{comment}
\usepackage{float}

\begin{document}

\pagestyle{fancy}
\rhead{\includegraphics[width=2.5cm]{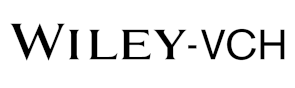}}

\title{Thermodynamic Multipoles and Dissipative Conductivities in Metallic Systems}
\maketitle

\author{Takumi Sato*}
\author{Satoru Hayami}

\begin{affiliations}
Graduate School of Science, Hokkaido University, Sapporo 060-0810, Japan\\
\end{affiliations}


\begin{abstract}
Multipoles provide a systematic framework for describing the electronic structures of quantum materials from a symmetry perspective.
Thermodynamic multipole moments in crystalline solids exhibit direct microscopic connections to certain allowed physical responses beyond symmetry; 
however, such relations have thus far been limited to  
dissipationless responses in equilibrium insulating systems. 
Here, this framework is extended at a heuristic level 
by focusing on the Fermi-surface contributions to thermodynamic multipole moments. 
These contributions establish direct relations to dissipative transport responses characteristic of metals, 
including charge and spin conductivities.
A key consequence is that the conductivities exhibit extrema, typically maxima, 
at chemical potentials where the corresponding Fermi-surface contributions to the multipoles vanish, 
specifically, the electric quadrupole for charge conductivity and the magnetic octupole for spin conductivity.
These findings uncover a previously overlooked aspect of thermodynamic multipole moments, opening  
a new perspective on dissipative transport in metallic systems.
\end{abstract}

\section{Introduction}
Multipole moments in solids have played important roles in the analysis of exotic quantum materials.
For example, once the ordered multipoles in a system are identified, 
the allowed physical responses can immediately be determined from symmetry arguments%
~\cite{kuramoto:ptp2008review, kusunose:jpsj2008, santini:rmp2009, kuramoto:jpsj2009review,
hayami:prb2018classification, watanabe:prb2018classification, yatsushiro:prb2021classification, hayami:jpsj2024review}.
Since the four types of multipoles classified according to their spacetime symmetry in a single atom as well as in cluster structures,
constitute a complete basis spanning the Hilbert space, any electronic degree of freedom in solids can, in principle, be described in terms of multipoles%
~\cite{hayami:jpsj2018, kusunose:jpsj2020, kusunose:prb2023, hayami:jpsj2024review, kuniyoshi:arxiv2026}.
However, such arguments do not go beyond symmetry analysis and only indicate the possible coexistence of ferroically ordered multipoles and the corresponding physical responses~\cite{suzuki:prb2017}.
Therefore, establishing microscopic relationships between multipole moments and physical responses 
remains challenging within symmetry analysis alone.

Important progress has been made for crystalline insulators.
According to the modern theory of multipole moments in crystalline solids, which overcomes the difficulty associated with the position operator in periodic systems, bulk multipole moments provide further insight: 
thermodynamic multipole moments are directly linked to some of the allowed physical responses.
For instance, the chemical potential derivatives of higher-rank electric and magnetic multipole moments in solids, 
including quadrupoles and higher-order moments, are directly related to 
the electric and magnetoelectric polarizabilities, respectively, in insulating systems at zero temperature%
~\cite{daido:prb2020, gao:prb2018spinMQM, shitade:prb2019spinMQM, shitade:prb2018orbitalMQM, gao:prb2018orbitalMQM,
oike:prb2025spinMO, TS:npjqmat2026, TS:arxiv2025orbitalMO}.
These responses are intrinsic, dissipationless, and valid for insulating systems.
There are, however, notable exceptions even in metals.
For example, the thermodynamic orbital magnetic quadrupole moment has been shown 
to be directly related to a nonlinear Hall response in metallic systems, which is finite only in metals while remaining intrinsic and dissipationless~\cite{gao:prb2018orbitalMQM}.
Nevertheless, dissipative responses such as the longitudinal conductivity, defined by $J_i=\sigma_{ij}E_j$ and $\sigma^{\mathrm{L}}_{ij}=(\sigma_{ij}+\sigma_{ji})/2$, are also believed to reflect the anisotropy of charge distributions in solids, just as the intrinsic Hall part, $\sigma^{\mathrm{H}}_{ij}=(\sigma_{ij}-\sigma_{ji})/2$, characterizes electronic structures with broken time-reversal symmetry.
Although the thermodynamic orbital magnetic dipole is known to be directly related to the Hall conductivity through the St\v{r}eda formula~\cite{streda:jphysc1982,widom:physletta1982,ceresoli:prb2006,xiao:rmp2010}, a corresponding microscopic relationship for 
dissipative (longitudinal) responses remains unclear.
Thus, the modern theory of multipoles still lacks a clear connection to dissipative transport phenomena.

In the present study, a heuristic but direct relationship is established 
between the intraband, or Fermi-surface, contributions to thermodynamic multipole moments and dissipative transport responses.
(The terminology ``intraband, or Fermi-surface, contributions'' is not strictly accurate; 
this point will be clarified in the following section.)
In particular, the Fermi-surface contribution to the thermodynamic electric quadrupole (EQ) moment~\cite{daido:prb2020} is shown to be directly related to the longitudinal charge conductivity.
This enables that the Fermi-surface contribution to the bulk EQ moment to be  
accessed through dissipative transport measurements, providing a contribution complementary to that obtained from the thermodynamic relation originating from interband terms.
Furthermore, direct relationships are discussed between the intraband, or Fermi-surface, contribution
to the thermodynamic spin magnetic octupole (MO) moment~\cite{oike:prb2025spinMO,TS:npjqmat2026} and the dissipative spin conductivity characteristic of time-reversal symmetry breaking collinear antiferromagnets with nonrelativistic spin splitting, such as altermagnets%
~\cite{solovyev:prb1997,sivadas:prl2016,noda:pccp2016,okugawa:jpcm2018,smejkal:sciadv2020,naka:natcomm2019,ahn:prb2019,
hayami:jpscp2020,hayami:jpsj2019spinsplit,hayami:prb2020spinsplit,yuan:prb2020,naka:prb2021perovskite,
yuan:prm2021,yuan:prb2021,hernandez:prl2021,smejkal:prx2022magnetoresistance,mazin:pnas2021,
smejkal:prx2022spingroup1,smejkal:prx2022spingroup2,
cheong:npj2025AMclassification,guo:advmat2025review,hu:prx2025}.
A key consequence is that the corresponding conductivities exhibit extrema, 
typically maxima when the Fermi-surface contributions to the thermodynamic multipole moments vanish.
These findings uncover a previously overlooked role of thermodynamic multipole moments 
and provide a new perspective on dissipative responses in metallic systems.

The present paper is organized as follows.
In Section~2, the expressions for thermodynamic EQ and spin MO moments 
are reviewed and compared with those for dissipative charge and spin conductivities, respectively.
In Section~3, model calculations are presented, and the relationships between thermodynamic multipole moments and conductivities are discussed.
The results are summarized in Section~4.
Throughout this paper, we use the units of $k_\mathrm{B} = c = \hbar = 1$
where $k_\mathrm{B}$ is the Boltzmann constant and $c$ is the speed of light
and $e<0$ is the charge of an electron.

\section{Thermodynamic Multipoles and Conductivities}
In this section, the expressions for the thermodynamic multipole moments and conductivity tensors are briefly 
reviewed, and their formal structures are compared.

The thermodynamic EQ $Q_{ij}$ is defined as~\cite{daido:prb2020}
\begin{equation}
  Q_{ij}  = -\frac{\partial F}{\partial [\partial_{r_j} E_i]} ,
\end{equation}
where $i,j=x,y,z$, $F$ is the free energy density, and 
$E_i = -\partial_{r_i} \phi (\bm{r})$ is the electric field, with $\phi (\bm{r})$ being the scalar potential.
The expression for Bloch electrons in periodic crystals is given by~\cite{daido:prb2020}
\begin{equation}
  Q_{ij} = e \sum_{n} \int \frac{d^d k}{(2\pi)^d} \left[
  g^{ij}_{n} f_n - G^{ij}_{n} \mathcal{G}_n -\frac{1}{12} \partial_{k_i} \partial_{k_j} \epsilon_n f_n' \right] 
  = Q_{ij}^{\mathrm{sea}} + Q_{ij}^{\mathrm{gdens}} + Q_{ij}^{\mathrm{surf}},
  \label{eq:EQ}
\end{equation}
where $d$ denotes the spatial dimension, and $\epsilon_n=\epsilon_{n\bm{k}}$ is 
the energy eigenvalue of the Bloch Hamiltonian $\hat{H}_{\bm{k}}$, 
satisfying $\hat{H}_{\bm{k}}\Ket{u_{n\bm{k}}} = \epsilon_n \Ket{u_{n\bm{k}}}$.
$f_n = [e^{(\epsilon_n-\mu)/T}+1]^{-1}$ and 
$\mathcal{G}_n = -T \ln \left(1 + e^{-(\epsilon_n-\mu)/T}\right)$ 
denote the Fermi distribution function and the grand potential density, respectively.
$g^{ij}_n = \mathrm{Re} \sum_{m}^{\neq n} A^{i}_{nm} A^{j}_{mn}$, 
which corresponds to the quantum metric~\cite{provost:cmp1980,berry:book1989,resta:epjb2011}, 
where $A^{i}_{nm}=\Braket{u_{n\bm{k}}|i\partial_{k_i}|u_{m\bm{k}}}$ denotes the interband Berry connection. 
$G^{ij}_n = 2\mathrm{Re}\sum_{m}^{\neq n} A^{i}_{nm} A^{j}_{mn} / (\epsilon_n - \epsilon_m)$, which is often referred to as 
the Berry connection polarizability (or positional shift)~\cite{gao:prl2014BCP}.
$Q_{ij}$ is a reducible rank-2 symmetric tensor with six independent components. 
It can be decomposed into its traceless and trace parts as
$Q^{Q_{2m}}_{ij} = Q_{ij} - Q_{kk} \delta_{ij}/3$ and $Q^{Q_0}_{ij} = Q_{kk} \delta_{ij}/3$,
where $Q^{Q_{2m}}_{ij}$ and $Q^{Q_0}_{ij}$ 
denote the traceless EQ and electric monopole components, respectively.
A brief derivation of Equation~\ref{eq:EQ} is provided in the Supporting Information~\cite{SI}.

The physical meaning of each term in Equation~\ref{eq:EQ} is briefly summarized 
as follows~\cite{daido:prb2020,onishi:nanolett2025}.
The first and second terms are associated with interband processes and therefore vanish in single-band systems, 
while they remain finite in both metals and insulators.
The first term $Q_{ij}^{\mathrm{sea}}$ represents contributions 
from electron wave packets~\cite{lapa:prb2019,gao:prl2019}, or, equivalently, 
from the spatial spread of Wannier functions within a unit cell~\cite{marzari:prb1997,marzari:rmp2012}.
The second term $Q_{ij}^{\mathrm{gdens}}$ originates from edge polarization in finite samples, 
but remains finite even in the thermodynamic limit~\cite{xiao:rmp2010}.
The third term $Q_{ij}^{\mathrm{surf}}$, which contains the inverse effective mass tensor 
$\partial_{k_i} \partial_{k_j} \epsilon_n$, originates from the Lindhard function for a free electron gas.
This term reflects the distortion of the band structure and, in particular, 
captures the deformation of the Fermi surface, since it remains finite only in metallic systems.

Before proceeding, we note that the decomposition into the grand-potential-density, Fermi-sea, 
and Fermi-surface contributions is not unique~\cite{karplus:pr1954AHE,haldane:prl2004,nagaosa:rmp2010AHE}.
For instance, part of the Fermi-sea contribution can be recast as a Fermi-surface contribution through partial integration.
In Reference~\cite{daido:prb2020}, this particular decomposition is adopted, as it provides a clear physical interpretation of each contribution.
It should also be noted that $Q_{ij}^{\mathrm{surf}}$ 
generally includes both intraband and interband contributions~\cite{gao:prb2015,resta:arxiv2017DWOAM,resta:jpcm2018DWSCW}.
Nevertheless, in single-band systems, this term can remain the only nonvanishing contribution to the thermodynamic EQ moment.

With these preliminaries in hand, we now present the central result of this work. 
We show that the Fermi-surface contribution to the thermodynamic EQ moment, 
$Q_{ij}^{\mathrm{surf}}$, is directly related to the dissipative longitudinal charge conductivity $\sigma^{\mathrm{L}}_{ij}$ as
\begin{equation}
  \sigma^{\mathrm{L}}_{ij} 
  = e^2 \tau \sum_{n} \int \frac{d^d k}{(2\pi)^d} \, \partial_{k_i} \partial_{k_j} \epsilon_n f_n 
  = 12 e \tau \int_{-\infty}^{\mu} d\mu' \, Q_{ij}^{\mathrm{surf}}(\mu'),
  \label{eq:Q_sigma}
\end{equation}
where $\tau$ is the phenomenological relaxation time~\cite{ashcroft,resta:epjb2011,resta:arxiv2017DWOAM,resta:jpcm2018DWSCW}.
A brief derivation of Equation~\ref{eq:Q_sigma} is provided in the Supporting Information~\cite{SI}.
Equation~\ref{eq:Q_sigma} remains valid even in metallic systems and at finite temperatures, 
indicating that dissipative transport is governed by the Fermi-surface contribution to the thermodynamic multipole moment. 
We note that this relation holds for both the anisotropic and isotropic parts of $\sigma^{\mathrm{L}}_{ij}$. 
This result should be contrasted with the well-established thermodynamic relation in insulating systems. 
In insulating systems at zero temperature, the following relation holds~\cite{daido:prb2020}:
\begin{equation}
  \chi_{ij} = -e \frac{\partial Q_{ij}}{\partial \mu} 
  = -e^2 \sum_{n}^{\mathrm{occ.}} \int \frac{d^d k}{(2\pi)^d} \, G^{ij}_n
  = -e \frac{\partial Q_{ij}^{\mathrm{gdens}}}{\partial \mu},
\end{equation}
where $\chi_{ij}$ denotes the electric susceptibility
that characterizes the polarization response to a static electric field.

Equation~\ref{eq:Q_sigma} may also be contrasted with the well-known connection between the thermodynamic orbital magnetic dipole and the anomalous Hall conductivity, which is valid for both insulating and metallic systems at finite temperatures:
\begin{equation}
  \sigma^{\mathrm{H}}_{ij} 
  = -e^2 \sum_n \int \frac{d^d k}{(2\pi)^d}\, \Omega_n^{ij} f_n 
  = e \epsilon_{ijk} \frac{\partial M_k^{\mathrm{gdens}}}{\partial \mu},
  \label{eq:M_sigma}
\end{equation}
where $M_k^{\mathrm{gdens}}$ denotes the grand-potential-density contribution to
the thermodynamic orbital magnetization $M_k = M_k^{\mathrm{sea}} + M_k^{\mathrm{gdens}}$%
~\cite{xiao:prl2005, xiao:prl2006thermoele, thonhauser:prl2005, ceresoli:prb2006, shi:prl2007, xiao:rmp2010, thonhauser:ijmpb2011},
and $\Omega_n^{ij}$ denotes the Berry curvature.
Equation~\ref{eq:M_sigma} reduces to the well-known St\v{r}eda formula%
~\cite{streda:jphysc1982,widom:physletta1982,ceresoli:prb2006,xiao:rmp2010}
in insulating systems at zero temperature.
Whereas the Hall conductivity is obtained from a chemical-potential derivative of a thermodynamic quantity, 
Equation~\ref{eq:Q_sigma} relates the dissipative conductivity to a chemical-potential integral of the Fermi-surface contribution.
The relation between the chemical-potential derivative of the orbital magnetization and the Hall conductivity can be understood in terms of the magnetization current. 
By contrast, the physical meaning of the relation between the chemical-potential integral of the EQ moment and the longitudinal conductivity is less transparent, and the present relation should therefore be regarded as heuristic.
Nevertheless, this contrast is of particular interest, as the latter is specific to dissipative responses.

A direct consequence of Equation~\ref{eq:Q_sigma} is that the conductivity exhibits 
an extremum---typically a maximum---when $Q_{ij}^{\mathrm{surf}}(\mu)$ vanishes.
This characteristic behavior provides a direct link between dissipative transport 
and the Fermi-surface contribution to the thermodynamic multipole moment.
Note that $Q_{ij}^{\mathrm{sea}}$ is also related to the dissipative conductivity over the entire frequency range through the Souza-Wilkens-Martin sum rule~\cite{souza:prb2000SWM_sumrule,resta:epjb2011,resta:arxiv2017DWOAM,onishi:prx2024}.
However, this connection is distinct from the above relations, as only static responses are considered here.

A similar relation can be established for the thermodynamic spin MO 
and the dissipative spin conductivity in collinear magnets without spin-orbit coupling.
The thermodynamic spin MO $M_{aij}$ is defined as
\begin{equation}
  M_{aij} = -\frac{\partial F}{\partial [\partial_{r_i} \partial_{r_j} B_a]} ,
\end{equation}
where $B_a$ denotes the Zeeman field.
For Bloch electrons in spin-conserving systems, in which the band off-diagonal components of spin operators vanish, the expression is given by~\cite{oike:prb2025spinMO, TS:npjqmat2026}
\begin{equation}
  M_{aij} = g \mu_{\mathrm{B}} \sum_{n} \int \frac{d^d k}{(2\pi)^d} \left[
  \sum_{m}^{\neq n} 
  \left\{ \frac{s^a_n+s^a_m}{2} \left( g^{ij}_{nm} f_n - G^{ij}_{nm} \mathcal{G}_n \right) \right\} 
  -\frac{1}{12} s^a_n \partial_{k_i} \partial_{k_j} \epsilon_n f_n'
  \right] 
  = M_{aij}^{\mathrm{sea}} + M_{aij}^{\mathrm{gdens}} + M_{aij}^{\mathrm{surf}},
  \label{eq:MO}
\end{equation}
where $g$ and $\mu_{\mathrm{B}}$ denote spin g-factor and Bohr magneton, respectively.
The quantities $g^{ij}_{nm}$ and $G^{ij}_{nm}$ denote the band-resolved contributions to $g^{ij}_{n}$ and $G^{ij}_{n}$, respectively, 
defined by $g^{ij}_{n} = \sum_{m}^{\neq n} g^{ij}_{nm}$ and $G^{ij}_{n} = \sum_{m}^{\neq n} G^{ij}_{nm}$, 
and $s^a_n = \braket{u_{n\bm{k}}|\hat{s}_a|u_{n\bm{k}}}$.
As in the case of $Q_{ij}$, $M_{aij}$ is also a reducible tensor with 18 independent components:
$M_{aij} = M_{aij}^{M_{3m}} + M_{aij}^{T_{2m}} + M_{aij}^{M_{1m}} + M_{aij}^{M'_{1m}}$, 
where each term represents, in order, the contributions from the totally symmetric MO, magnetic toroidal quadrupole, 
magnetic dipole, and anisotropic magnetic dipole~\cite{TS:npjqmat2026}.
A brief derivation of Equation~\ref{eq:MO} is provided in the Supporting Information~\cite{SI}.
In spin-conserving systems, the expression for the spin MO is obtained from that for the EQ by replacing charge with spin. 
Accordingly, each term in Equation~\ref{eq:MO} can be understood 
as the spin analogue of the corresponding term in Equation~\ref{eq:EQ}. 
More specifically, $M_{aij}^{\mathrm{sea}}$ represents the wave packet contribution to the spin MO~\cite{tahir:prl2023}, 
$M_{aij}^{\mathrm{gdens}}$ corresponds to the surface spin magnetic quadrupole, 
and $M_{aij}^{\mathrm{surf}}$, which survives only in metallic systems, 
originates from the band dispersion and plays a central role in the following discussion.

In insulating systems at zero temperature, the following thermodynamic relation holds~\cite{oike:prb2025spinMO,TS:npjqmat2026}:
\begin{equation}
  \alpha_{aij} = e \frac{\partial M_{aij}}{\partial \mu} 
  = e g \mu_{\mathrm{B}} \sum_{n}^{\mathrm{occ.}} \int \frac{d^d k}{(2\pi)^d}
  \sum_{m}^{\neq n} \frac{s^a_n+s^a_m}{2} G^{ij}_{nm} 
  = e \frac{\partial M_{aij}^{\mathrm{gdens}}}{\partial \mu},
\end{equation}
where $\alpha_{aij}$ denotes the quadrupolar magnetoelectric susceptibility characterizing 
the polarization (magnetization) response to a magnetic (an electric) field gradient~\cite{shitade:prb2025}. 
In parallel with Equation~\ref{eq:Q_sigma}, a heuristic relation is 
found between the Fermi-surface contribution $M_{aij}^{\mathrm{surf}}$ and the dissipative spin conductivity, 
which remains valid even in metallic systems and at finite temperatures, although 
only in spin-conserving systems:
\begin{equation}
  \sigma_{aij} 
  = e \tau \sum_{n} \int \frac{d^d k}{(2\pi)^d} \, s^a_n \partial_{k_i} \partial_{k_j} \epsilon_n f_n 
  = \frac{12e\tau}{g\mu_{\mathrm{B}}} \int_{-\infty}^{\mu} d\mu' \, M_{aij}^{\mathrm{surf}}(\mu'),
  \label{eq:MO_sigma}
\end{equation}
where $\sigma_{aij}$ characterizes the spin current $\hat{J}_{ia} = \left\{ \hat{v}_i, \hat{s}_a \right\}/2$ 
response to an electric field: $J_{ia} = \sigma_{aij} E_j$~\cite{sinova:prl2004,mook:prr2020}. 
Equation~\ref{eq:MO_sigma} is valid only when the band-off-diagonal components of the spin operator can be neglected.
A derivation of this equation is provided in the Supporting Information~\cite{SI}. 
Analogous to the relation between $Q_{ij}^{\mathrm{surf}}$ and $\sigma^{\mathrm{L}}_{ij}$ in Equation~\ref{eq:Q_sigma}, 
this relation shows that the dissipative spin conductivity is governed by 
the Fermi-surface contribution to the thermodynamic spin MO. 
As a direct consequence, the spin conductivity exhibits an extremum when $M_{aij}^{\mathrm{surf}}(\mu)$ vanishes.
We also note that $\sigma_{aij}$ has the same 18 independent components as $M_{aij}$ 
and can therefore be decomposed in the same manner. 

The expressions summarized in this section demonstrate that thermodynamic multipole moments and conductivity tensors share closely related structures.  
Whereas the interband terms are connected to intrinsic equilibrium responses, 
the intraband terms, or Fermi-surface terms, more precisely those arising from the band dispersion, 
are associated with dissipative transport responses through the heuristic relations 
in Equation~\ref{eq:Q_sigma} and Equation~\ref{eq:MO_sigma}. 
In particular, these relations imply that the conductivities exhibit extrema 
when the corresponding Fermi-surface multipole contributions vanish. 
To elucidate these features, numerical results are presented in the following section.

\section{Model Calculations and Results}
In this section, the formal results obtained in the previous section are examined 
through numerical calculations for a representative model.
We consider the following Hamiltonian for a spin-orbit-coupling-free metallic
altermagnet in the rutile structure~\cite{roig:prb2024,antonenko:prl2025}, which provides 
a minimal spin-conserving model suitable for discussing finite charge and spin conductivities:
\begin{equation}
    \hat{H}_{\bm{k}}
    = \varepsilon_{0,\bm{k}} + t_{x,\bm{k}} \hat{\tau}_{x} + t_{z,\bm{k}} \hat{\tau}_{z} + J \hat{\tau}_{z} \hat{\sigma}_{z} ,
\end{equation}
where $\hat{\tau}_i$ and $\hat{\sigma}_i$ denote the Pauli matrices 
acting in the sublattice and spin spaces, respectively. 
$t_{x,\bm{k}}$ and $t_{z,\bm{k}}$ represents the inter- and intra-sublattice hoppings, respectively, and
$J$ denotes the magnitude of the magnetic moment localized at each sublattice. 
The coefficients entering the Hamiltonian are given by
\begin{equation}
    \varepsilon_{0,\bm{k}} =
    t_1 ( \cos k_x + \cos k_y ) - \mu + t_2 \cos k_z + t_3 \cos k_x \cos k_y 
    +t_4 ( \cos k_x + \cos k_y ) \cos k_z + t_5 \cos k_x \cos k_y \cos k_z ,
\end{equation}
\begin{equation}
    t_{x,\bm{k}}
    = t_8 \cos \frac{k_x}{2} \cos \frac{k_y}{2} \cos \frac{k_z}{2} , 
\end{equation}
\begin{equation}
    t_{z,\bm{k}}
    = t_6 \sin k_x \sin k_y + t_7 \sin k_x \sin k_y \cos k_z .
\end{equation}
The hopping parameters are chosen as
$t_1=0$, $t_2=0.13$, $t_3=0$, $t_4=-0.02$, $t_5=0.015$,
$t_6=0$, $t_7=0.03$, $t_8=0.33$, and $\mu=-0.01$ 
to reproduce the nonmagnetic band structure of MnF$_2$~\cite{roig:prb2024}.
In the numerical simulations, $J$ and the temperature are set to 0.1 and 0.01, respectively. 
The Hamiltonian preserves the mirror symmetries $\mathcal{M}_x$, $\mathcal{M}_y$, and $\mathcal{M}_z$, 
as well as the antiunitary symmetry $C_{4z} \mathcal{T}$, 
where $C_{4z}$ and $\mathcal{T}$ denote a fourfold rotation around the $z$-axis and the time-reversal operation, respectively.
These symmetry constraints restrict the allowed EQ components to $Q_{zz}$ and $Q_{xx}=Q_{yy}$. 
For the spin MO, the allowed components are $M_{zxy}$, and $M_{xyz} = M_{yzx}$~\cite{xiao:prl2022hkspaceBCP}.
In the present spin-conserving model, however, only $M_{zxy}$ remains nonvanishing.

\begin{figure}
  \includegraphics[width=\linewidth]{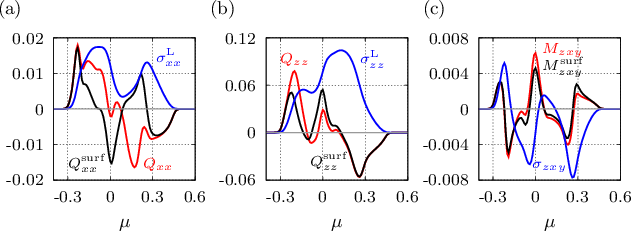}
  \caption{
  Chemical-potential dependence of the Fermi-surface contribution (black), 
  total thermodynamic multipole (red), and the corresponding dissipative conductivity tensor (blue) are shown.
  In each panel, the thermodynamic multipole and conductivity are plotted 
  in arbitrary units with different vertical scales.
  }
  \label{fig:mudep}
\end{figure}

Figure~\ref{fig:mudep}(a) [(b)] shows the chemical-potential dependence of 
 $Q^{\mathrm{surf}}_{xx}$, $Q_{xx}$, and $\sigma^{\mathrm{L}}_{xx} \propto \int d \mu\, Q^{\mathrm{surf}}_{xx}$ 
[$Q^{\mathrm{surf}}_{zz}$, $Q_{zz}$, and $\sigma^{\mathrm{L}}_{zz} \propto \int d \mu\, Q^{\mathrm{surf}}_{zz}$]. 
In (c), the chemical-potential dependence of 
$M^{\mathrm{surf}}_{zxy}$, $M_{zxy}$, and $\sigma_{zxy} \propto \int d \mu\, M^{\mathrm{surf}}_{zxy}$ 
is shown.
The zero crossings of the Fermi-surface contribution (black curve) 
coincide with extrema of the conductivity (blue curve), 
and in many cases correspond to its local maxima. 
To put it the other way around, the conductivities 
reach their global maximum at the chemical potential where the Fermi-surface contribution vanishes.
Because the thermodynamic multipoles also include Fermi-sea and grand-potential-density contributions, 
a one-to-one correspondence between the total multipole and the conductivity is not always observed, as seen in panel (a).
The interband contributions, $Q_{ij}^{\mathrm{sea}}$ and $Q_{ij}^{\mathrm{gdens}}$, 
and consequently the quantum geometry, would play an important role in this case. 
Meanwhile, a clear correlation emerges in panels (b) and (c), 
where the conductivity tends to reach a local maximum when the thermodynamic multipole becomes nearly zero. 
We again note that the Fermi-surface term vanishes in all cases when the conductivities reach their extrema. 

This behavior is counterintuitive and therefore of particular interest.
Although thermodynamic multipole moments often vanish at specific parameter values%
~\cite{daido:prb2020, gao:prb2018spinMQM, shitade:prb2019spinMQM, shitade:prb2018orbitalMQM, gao:prb2018orbitalMQM,
oike:prb2025spinMO, TS:npjqmat2026, TS:arxiv2025orbitalMO}, 
the present results imply that such zeros should not be simply interpreted 
as the disappearance of the corresponding multipole order. 
The conductivity exhibits an extremum, typically a maximum, 
when the corresponding Fermi-surface contribution vanishes. 
Therefore, in some cases, a thermodynamic multipole becomes small or even vanishes in a parameter region 
where the dissipative response characteristic of the corresponding multipole order is enhanced, 
as seen in the present numerical results.
Since the total thermodynamic multipoles also contain the Fermi-sea and grand-potential-density contributions, 
this correspondence is not an exact one-to-one relation 
between the zeros of the total multipoles and the extrema of the conductivities. 
Nevertheless, it is intriguing that such a correspondence can approximately emerge 
when the Fermi-surface contribution plays a dominant role. 
Thus, these results indicate that not only the values of the thermodynamic multipoles themselves, 
but also their decomposition and chemical-potential dependence, 
play an essential role in characterizing and identifying multipole orders.
These findings thus provide a new perspective on the relationship 
between the thermodynamic multipoles and dissipative transport in metallic systems.

\section{Conclusion}
In this study, direct relationships between the Fermi-surface contributions to thermodynamic multipole moments 
and dissipative conductivities have been established at a heuristic level. 
The Fermi-surface contributions to the thermodynamic EQ and, 
in spin-conserving systems such as altermagnets, to the thermodynamic spin MO 
are directly related to the longitudinal charge and spin conductivities, respectively.
A key consequence of these relations is that the corresponding conductivity 
exhibits extrema, typically maxima, at chemical potentials where 
the associated Fermi-surface contributions to the thermodynamic multipole vanish.
This result indicates that the vanishing of a thermodynamic multipole does not necessarily imply 
the disappearance of the corresponding multipole order.
Rather, such points can be accompanied by enhanced dissipative transport responses characteristic of that order.
These results reveal a previously overlooked aspect of thermodynamic multipole moments 
and suggest that not only the values of the multipoles themselves, but also their chemical-potential dependence, are important for discussing multipole orders.

\medskip
\textbf{Acknowledgements} \par 
This research was supported by JSPS KAKENHI Grants Numbers JP22H00101, JP22H01183, JP23H04869, JP23K03288, and by JST CREST (JPMJCR23O4) and JST FOREST (JPMJFR2366).

\medskip


\clearpage
\begin{center}
\textbf{\large Supporting Information for\\
``Thermodynamic Multipoles and Dissipative Conductivities in Metallic Systems''}\\
\vspace{1em}
{Takumi Sato and Satoru Hayami}\\
\textit{Graduate School of Science, Hokkaido University, Sapporo 060-0810, Japan}\\
\end{center}
\vspace{0.5cm}

\setcounter{section}{0}
\setcounter{equation}{0}
\setcounter{figure}{0}
\setcounter{table}{0}
\renewcommand{\thesection}{S\arabic{section}}
\renewcommand{\theequation}{S\arabic{equation}}
\renewcommand{\thefigure}{S\arabic{figure}}
\renewcommand{\thetable}{S\arabic{table}}

\section{Derivation of Thermodynamic Multipoles}
\label{sec:derivation_MP}
In this section, we briefly summarize the derivations of the thermodynamic electric quadrupole (EQ) and spin magnetic octupole (MO)
using the Kubo formula~\cite{daido:prb2020,oike:prb2025spinMO,TS:npjqmat2026}.

Our starting point is the differential form of the free-energy density $F(\bm{r})$ 
in crystalline solids with a slowly varying 
electric and Zeeman fields, $\bm{E}(\bm{r}) = -\nabla \phi (\bm{r})$ and $\bm{B}(\bm{r})$:
\begin{equation}
    dF = -SdT - Nd\mu - Q_{ij} d[\partial_{r_j} E_{i}] - M_{aij} d[\partial_{r_i} \partial_{r_j} B_{a}] ,
    \label{eq:differential_FE}
\end{equation}
where $S$, $T$, $N$, and $\mu$ are the entropy, temperature, particle number, and chemical potential, respectively.
Here, $Q_{ij}$ and $M_{aij}$ $(a,i,j = x, y, z)$ denote the EQ and spin MO, respectively.
In the following, we retain only the terms involving 
$\partial_{r_j} E_{i}$ and $\partial_{r_i} \partial_{r_j} B_{a}$.
Using the Maxwell relations derived from Equation~\ref{eq:differential_FE}, 
the particle-number change is related to the multipole moment as 
\begin{equation}
  \frac{\partial N}{\partial [-\partial_{r_i} \partial_{r_j} \phi]} = \frac{\partial Q_{ij}}{\partial \mu}, \quad 
  \frac{\partial N}{\partial [ \partial_{r_i} \partial_{r_j} B_a]} = \frac{\partial M_{aij}}{\partial \mu} .
  \label{eq:Maxwell}
\end{equation}

The particle-number changes in Equation~\ref{eq:Maxwell} can be computed from the following correlation functions:
\begin{equation}
  \chi_{N \rho} (\bm{q},\omega) =
    -e\sum_{nm} \int \frac{d^d k}{(2\pi)^d}
    |\braket{u_{n\bm{k}_{-}}|u_{m\bm{k}_{+}}}|^2
    \frac{f(\epsilon_{n\bm{k}_{-}}-\mu)-f(\epsilon_{m\bm{k}_{+}}-\mu)}{\omega+i\delta + \epsilon_{n\bm{k}_{-}} -\epsilon_{m\bm{k}_{+}}} ,
    \label{eq:chi_nrho} 
\end{equation}
and
\begin{equation}
  \chi_{N M_a} (\bm{q},\omega) =
    -\sum_{nm} \int \frac{d^d k}{(2\pi)^d}
    \braket{u_{n\bm{k}_{-}}|u_{m\bm{k}_{+}}}
    \braket{u_{m\bm{k}_{+}}|g\mu_{\mathrm{B}}\hat{s}_a|u_{n\bm{k}_{-}}}
    \frac{f(\epsilon_{n\bm{k}_{-}}-\mu)-f(\epsilon_{m\bm{k}_{+}}-\mu)}{\omega+i\delta + \epsilon_{n\bm{k}_{-}} -\epsilon_{m\bm{k}_{+}}} 
    \label{eq:chi_nM} ,
\end{equation}
which describe the linear responses
\begin{equation}
  \delta \langle N \rangle (\bm{q},\omega) = -\chi_{N \rho} (\bm{q},\omega) \phi (\bm{q},\omega) ,
\end{equation} 
and
\begin{equation}
  \delta \langle N \rangle (\bm{q},\omega) = \chi_{N M_a} (\bm{q},\omega) B_a (\bm{q},\omega) ,
\end{equation}
respectively.
Here, we use the notation 
\begin{equation}
    \hat{H} \Ket{\psi_{n\bm{k}}} = \epsilon_{n\bm{k}} \Ket{\psi_{n\bm{k}}} , \quad
    \Ket{u_{n\bm{k}}} = e^{-i\bm{k}\cdot \hat{\bm{r}}} \Ket{\psi_{n\bm{k}}} ,\quad
    \bm{k}_{\pm} = \bm{k} \pm \frac{\bm{q}}{2} , \quad
    f(z) = [e^{\beta z}+1]^{-1} ,
    \label{eq:notations}
\end{equation}
where $\hat{H}$ is the single-electron Hamiltonian in a periodic crystal and 
$\hat{H} ({\bm{k}}) = e^{-i\bm{k}\cdot \hat{\bm{r}}} \hat{H} e^{i\bm{k}\cdot \hat{\bm{r}}}$ is the Bloch Hamiltonian.

The EQ and spin MO are obtained from
\begin{equation}
  Q_{ij}(\mu) = \int_{-\infty}^{\mu} d\mu' \, \frac{\partial Q_{ij}(\mu')}{\partial \mu'} , \quad
  \frac{\partial Q_{ij}(\mu)}{\partial \mu}
  = \lim_{\bm{q} \to \bm{0}}
  \frac{-1}{2} \partial_{q_i}\partial_{q_j} \lim_{\delta \to +0} \chi_{N \rho} (\bm{q},0) ,
\end{equation}
and
\begin{equation}
  M_{aij}(\mu) = \int_{-\infty}^{\mu} d\mu' \, \frac{\partial M_{aij}(\mu')}{\partial \mu'} , \quad
  \frac{\partial M_{aij}(\mu)}{\partial \mu}
  = \lim_{\bm{q} \to \bm{0}}
  \frac{-1}{2} \partial_{q_i}\partial_{q_j} \lim_{\delta \to +0} \chi_{N M_a} (\bm{q},0) ,
\end{equation}
respectively.

For the EQ, we obtain~\cite{daido:prb2020}
\begin{equation}
  Q_{ij} = e \sum_{n} \int \frac{d^d k}{(2\pi)^d} \left[
  g^{ij}_{n} f_n - G^{ij}_{n} \mathcal{G}_n -\frac{1}{12} \partial_{k_i} \partial_{k_j} \epsilon_n f_n' \right] ,
\end{equation}
where
\begin{equation}
  g^{ij}_n = \sum_{m}^{\neq n} g^{ij}_{nm} = \mathrm{Re} \sum_{m}^{\neq n} A^{i}_{nm} A^{j}_{mn} ,\quad
  G^{ij}_n = \sum_{m}^{\neq n} G^{ij}_{nm} = 2\mathrm{Re}\sum_{m}^{\neq n} \frac{A^{i}_{nm} A^{j}_{mn}} {\epsilon_n - \epsilon_m} ,
\end{equation}
and $A^{i}_{nm}=\Braket{u_{n\bm{k}}|i\partial_{k_i}|u_{m\bm{k}}}$ is the interband Berry connection. 
Here, $\mathcal{G}_n = -T \ln \left(1 + e^{-(\epsilon_n-\mu)/T}\right)$ denotes the grand-potential density.

For the spin MO in spin-conserving systems, where the band-off-diagonal components of the spin operators vanish, 
we obtain~\cite{oike:prb2025spinMO, TS:npjqmat2026}
\begin{equation}
  M_{aij} = g \mu_{\mathrm{B}} \sum_{n} \int \frac{d^d k}{(2\pi)^d} \left[
  \sum_{m}^{\neq n} 
  \left\{ \frac{s^a_n+s^a_m}{2} \left( g^{ij}_{nm} f_n - G^{ij}_{nm} \mathcal{G}_n \right) \right\} 
  -\frac{1}{12} s^a_n \partial_{k_i} \partial_{k_j} \epsilon_n f_n'
  \right] ,
\end{equation}
where $s^a_n = \braket{u_{n\bm{k}}|\hat{s}_a|u_{n\bm{k}}}$.

\section{Derivation of Conductivities}
\label{sec:derivation_sigma}
We next derive the dissipative charge and spin conductivities using the Kubo formula 
by introducing the phenomenological relaxation time $\tau$%
~\cite{ashcroft,resta:epjb2011,resta:arxiv2017DWOAM,resta:jpcm2018DWSCW,sinova:prl2004,mook:prr2020}.

We define the paramagnetic charge and spin current operators as
\begin{equation}
  \hat{J}_i = e \hat{v}_i ,\quad 
  \hat{J}_{ia} = \frac{1}{2} \{ \hat{v}_i, \hat{s}_a\} , 
\end{equation}
respectively.
Here, $\hat{v}_i = i[\hat{H}, \hat{r}_i]$ and $\hat{s}_a$ are the velocity and spin operators, respectively.
The responses of interest here are 
\begin{equation}
  J_i    (\bm{q},\omega) = \sigma_{ij}  (\bm{q},\omega) E_j (\bm{q},\omega) ,\quad
  J_{ia} (\bm{q},\omega) = \sigma_{aij} (\bm{q},\omega) E_j (\bm{q},\omega) ,
\end{equation} 
where $E_j (\bm{q},\omega) = i(\omega + i\delta) A_j (\bm{q},\omega)$ is the electric field. 
The corresponding response tensors are written as 
\begin{equation}
  \sigma_{ij}  (\bm{q},\omega) = \frac{\chi_{J_i    J_j}(\bm{q},\omega) - D_{ij}} {i(\omega + i\delta)} ,\quad
  \sigma_{aij} (\bm{q},\omega) = \frac{\chi_{J_{ia} J_j}(\bm{q},\omega) - D_{aij}}{i(\omega + i\delta)} ,
\end{equation}
where $\chi_{J_i J_j}$ and $\chi_{J_{ia} J_j}$ are the current-current and spin-current-current correlation functions, 
respectively, defined through 
\begin{equation}
  \delta \langle J_i    \rangle (\bm{q},\omega) = \chi_{J_i    J_j} (\bm{q},\omega) A_j (\bm{q},\omega) , \quad
  \delta \langle J_{ia} \rangle (\bm{q},\omega) = \chi_{J_{ia} J_j} (\bm{q},\omega) A_j (\bm{q},\omega) .
\end{equation} 
The correlation functions are expressed as
\begin{equation}
  \begin{array}{c} \displaystyle
      \chi_{J_i J_j}(\bm{q},\omega) =
      \frac{-e^2}{4}\sum_{nm} \int \frac{d^d k}{(2\pi)^d}
      \braket{u_{n\bm{k}_{-}}| \hat{v}_i(\bm{k}_+) + \hat{v}_i(\bm{k}_-) |u_{m\bm{k}_{+}}}
      \braket{u_{m\bm{k}_{+}}| \hat{v}_j(\bm{k}_+) + \hat{v}_j(\bm{k}_-) |u_{n\bm{k}_{-}}} \\ \displaystyle
      \times
      \frac{f(\epsilon_{n\bm{k}_{-}}-\mu)-f(\epsilon_{m\bm{k}_{+}}-\mu)}{\omega+i\delta + \epsilon_{n\bm{k}_{-}} -\epsilon_{m\bm{k}_{+}}} 
    \end{array}
    \label{eq:chi_JiJj} ,
\end{equation}
\begin{equation}
  \begin{array}{c} \displaystyle
      \chi_{J_{ia} J_j}(\bm{q},\omega) =
      \frac{-e}{8}\sum_{nm} \int \frac{d^d k}{(2\pi)^d}
      \braket{u_{n\bm{k}_{-}}| \{ \hat{v}_i(\bm{k}_+) + \hat{v}_i(\bm{k}_-), \hat{s}_a \} |u_{m\bm{k}_{+}}}
      \braket{u_{m\bm{k}_{+}}| \hat{v}_j(\bm{k}_+) + \hat{v}_j(\bm{k}_-) |u_{n\bm{k}_{-}}} \\ \displaystyle
      \times
      \frac{f(\epsilon_{n\bm{k}_{-}}-\mu)-f(\epsilon_{m\bm{k}_{+}}-\mu)}{\omega+i\delta + \epsilon_{n\bm{k}_{-}} -\epsilon_{m\bm{k}_{+}}} 
    \end{array}
    \label{eq:chi_JiaJj} .
\end{equation}
The diamagnetic contributions can be expressed as 
\begin{equation}
  D_{ij}  = \chi_{J_i    J_j}(\bm{q} \to \bm{0},\omega=0) ,\quad 
  D_{aij} = \chi_{J_{ia} J_j}(\bm{q} \to \bm{0},\omega=0) . 
\end{equation}

Here, we are interested in the uniform conductivities: 
$\sigma_{(a)ij} (\omega) = \sigma_{(a)ij} (\bm{q} = \bm{0}, \omega)
=\sigma^{\mathrm{L}}_{(a)ij} (\omega) + \sigma^{\mathrm{H}}_{(a)ij} (\omega)$.
The longitudinal parts are expressed as
\begin{equation}
  \sigma^{\mathrm{L}}_{ij} (\omega) =
  \frac{\chi_{J_i J_j}(\bm{0},\omega \to 0) - D_{ij}}{i(\omega+i\delta)} =
  \frac{e^2}{i(\omega+i\delta)} \sum_n \int \frac{d^d k}{(2\pi)^d} 
  \partial_{k_i} \epsilon_n \partial_{k_j} \epsilon_n f_n' ,
\end{equation}
\begin{equation}
  \sigma^{\mathrm{L}}_{aij} (\omega) =
  \frac{\chi_{J_{ia} J_j}(\bm{0},\omega \to 0) - D_{aij}}{i(\omega+i\delta)} =
  \frac{e}{i(\omega+i\delta)} \sum_n \int \frac{d^d k}{(2\pi)^d} 
  \braket{u_{n\bm{k}}|\frac{1}{2}\{ \hat{v}_i(\bm{k}) , \hat{s}_a \}|u_{n\bm{k}}} \partial_{k_j} \epsilon_n f_n' .
\end{equation}
In the static limit, we obtain
\begin{equation}
  \sigma^{\mathrm{L}}_{ij} (0) =
  -e^2 \tau \sum_n \int \frac{d^d k}{(2\pi)^d} 
  \partial_{k_i} \epsilon_n \partial_{k_j} \epsilon_n f_n' ,
\end{equation}
\begin{equation}
  \sigma^{\mathrm{L}}_{aij} (0) =
  -e \tau \sum_n \int \frac{d^d k}{(2\pi)^d} 
  \braket{u_{n\bm{k}}|\frac{1}{2}\{ \hat{v}_i(\bm{k}) , \hat{s}_a \}|u_{n\bm{k}}} \partial_{k_j} \epsilon_n f_n' ,
\end{equation}
where we introduced the phenomenological relaxation time $\tau = 1/\delta$.
By performing integration by parts and imposing $\braket{u_{n\bm{k}}|\hat{s}_a|u_{m\bm{k}}}=0$ for $n \neq m$, 
we obtain Equations~(3) and (9) in the main text.

The Hall parts are expressed as
\begin{equation}
  \sigma^{\mathrm{H}}_{ij} (\omega) =
  \frac{\chi_{J_i J_j}(\bm{0},\omega) - \chi_{J_i J_j}(\bm{0},\omega \to 0)}{i(\omega+i\delta)} =
  -ie^2 \sum_n \sum_{m}^{\neq n} \int \frac{d^d k}{(2\pi)^d} 
  \frac{\braket{u_{n\bm{k}}|\hat{v}_i(\bm{k})|u_{m\bm{k}}} \braket{u_{m\bm{k}}|\hat{v}_j(\bm{k})|u_{n\bm{k}}}}
  {\omega + i\delta + \epsilon_n-\epsilon_m}
  \frac{f_n - f_m}{\epsilon_n-\epsilon_m} ,
\end{equation}
\begin{equation}
  \sigma^{\mathrm{H}}_{aij} (\omega) =
  \frac{\chi_{J_{ia} J_j}(\bm{0},\omega) - \chi_{J_{ia} J_j}(\bm{0},\omega \to 0)}{i(\omega+i\delta)} =
  -ie \sum_n \sum_{m}^{\neq n} \int \frac{d^d k}{(2\pi)^d} 
  \frac{\braket{u_{n\bm{k}}|\frac{1}{2}\{ \hat{v}_i(\bm{k}) , \hat{s}_a \}|u_{m\bm{k}}} 
  \braket{u_{m\bm{k}}|\hat{v}_j(\bm{k})|u_{n\bm{k}}}}
  {\omega + i\delta + \epsilon_n-\epsilon_m}
  \frac{f_n - f_m}{\epsilon_n-\epsilon_m} .
\end{equation}
In the static limit, we obtain the well-known results:
\begin{equation}
  \sigma^{\mathrm{H}}_{ij} (0) = 
  -e^2 \sum_n \int \frac{d^d k}{(2\pi)^d} \Omega_n^{ij} f_n ,\quad
  \sigma^{\mathrm{H}}_{aij} (0) = 
  -e \sum_n \int \frac{d^d k}{(2\pi)^d} \Omega_n^{aij} f_n ,
\end{equation}
where $\Omega_n^{ij}$ and $\Omega_n^{aij}$ denote the Berry curvature and spin Berry curvature, 
respectively, defined as 
\begin{equation}
  \Omega_n^{ij} = 
  -2\mathrm{Im} \sum_{m}^{\neq n} 
  \frac{\braket{u_{n\bm{k}}|\hat{v}_i(\bm{k})|u_{m\bm{k}}}\braket{u_{m\bm{k}}|\hat{v}_j(\bm{k})|u_{n\bm{k}}}}
  {(\epsilon_n-\epsilon_m)^2} ,\quad
  \Omega_n^{aij} =
  -2\mathrm{Im} \sum_{m}^{\neq n} 
  \frac{\braket{u_{n\bm{k}}|\frac{1}{2}\{ \hat{v}_i(\bm{k}) , \hat{s}_a \}|u_{m\bm{k}}} 
  \braket{u_{m\bm{k}}|\hat{v}_j(\bm{k})|u_{n\bm{k}}}}
  {(\epsilon_n-\epsilon_m)^2} .
\end{equation}


\begin{thebibliography}{10}
\providecommand{\url}[1]{\texttt{#1}}
\providecommand{\urlprefix}{URL }

\bibitem{kuramoto:ptp2008review}
Y.~Kuramoto,
\newblock \emph{Progress of Theoretical Physics Supplement} \textbf{2008},
  \emph{176} 77.

\bibitem{kusunose:jpsj2008}
H.~Kusunose,
\newblock \emph{Journal of the Physical Society of Japan} \textbf{2008},
  \emph{77}, 6 064710.

\bibitem{santini:rmp2009}
P.~Santini, S.~Carretta, G.~Amoretti, R.~Caciuffo, N.~Magnani, G.~H. Lander,
\newblock \emph{Rev. Mod. Phys.} \textbf{2009}, \emph{81} 807.

\bibitem{kuramoto:jpsj2009review}
Y.~Kuramoto, H.~Kusunose, A.~Kiss,
\newblock \emph{Journal of the Physical Society of Japan} \textbf{2009},
  \emph{78}, 7 072001.

\bibitem{hayami:prb2018classification}
S.~Hayami, M.~Yatsushiro, Y.~Yanagi, H.~Kusunose,
\newblock \emph{Phys. Rev. B} \textbf{2018}, \emph{98} 165110.

\bibitem{watanabe:prb2018classification}
H.~Watanabe, Y.~Yanase,
\newblock \emph{Phys. Rev. B} \textbf{2018}, \emph{98} 245129.

\bibitem{yatsushiro:prb2021classification}
M.~Yatsushiro, H.~Kusunose, S.~Hayami,
\newblock \emph{Phys. Rev. B} \textbf{2021}, \emph{104} 054412.

\bibitem{hayami:jpsj2024review}
S.~Hayami, H.~Kusunose,
\newblock \emph{Journal of the Physical Society of Japan} \textbf{2024},
  \emph{93}, 7 072001.

\bibitem{hayami:jpsj2018}
S.~Hayami, H.~Kusunose,
\newblock \emph{Journal of the Physical Society of Japan} \textbf{2018},
  \emph{87}, 3 033709.

\bibitem{kusunose:jpsj2020}
H.~Kusunose, R.~Oiwa, S.~Hayami,
\newblock \emph{Journal of the Physical Society of Japan} \textbf{2020},
  \emph{89}, 10 104704.

\bibitem{kusunose:prb2023}
H.~Kusunose, R.~Oiwa, S.~Hayami,
\newblock \emph{Phys. Rev. B} \textbf{2023}, \emph{107} 195118.

\bibitem{kuniyoshi:arxiv2026}
S.~Kuniyoshi, R.~Oiwa, S.~Hayami,
\newblock {Theory of Many-Body Multipole Operators in Single-Centered Electron
  Systems: Two-Body Toroidal Monopoles in Spinless Orbitals}, \textbf{2026},
\newblock \urlprefix\url{https://arxiv.org/abs/2603.10620}.

\bibitem{suzuki:prb2017}
M.-T. Suzuki, T.~Koretsune, M.~Ochi, R.~Arita,
\newblock \emph{Phys. Rev. B} \textbf{2017}, \emph{95} 094406.

\bibitem{daido:prb2020}
A.~Daido, A.~Shitade, Y.~Yanase,
\newblock \emph{Phys. Rev. B} \textbf{2020}, \emph{102} 235149.

\bibitem{gao:prb2018spinMQM}
Y.~Gao, D.~Vanderbilt, D.~Xiao,
\newblock \emph{Phys. Rev. B} \textbf{2018}, \emph{97} 134423.

\bibitem{shitade:prb2019spinMQM}
A.~Shitade, A.~Daido, Y.~Yanase,
\newblock \emph{Phys. Rev. B} \textbf{2019}, \emph{99} 024404.

\bibitem{shitade:prb2018orbitalMQM}
A.~Shitade, H.~Watanabe, Y.~Yanase,
\newblock \emph{Phys. Rev. B} \textbf{2018}, \emph{98} 020407.

\bibitem{gao:prb2018orbitalMQM}
Y.~Gao, D.~Xiao,
\newblock \emph{Phys. Rev. B} \textbf{2018}, \emph{98} 060402.

\bibitem{oike:prb2025spinMO}
J.~\ifmmode~\bar{O}\else \={O}\fi{}ik\'e, R.~Peters, K.~Shinada,
\newblock \emph{Phys. Rev. B} \textbf{2025}, \emph{112} 134412.

\bibitem{TS:npjqmat2026}
T.~Sato, S.~Hayami,
\newblock \emph{npj Quantum Materials} \textbf{2026}, \emph{11} 32.

\bibitem{TS:arxiv2025orbitalMO}
T.~Sato, S.~Hayami,
\newblock {Orbital magnetic octupole in crystalline solids and characterization
  of orbital altermagnetism}, \textbf{2026},
\newblock \urlprefix\url{https://arxiv.org/abs/2512.24269}.

\bibitem{streda:jphysc1982}
P.~Streda,
\newblock \emph{Journal of Physics C: Solid State Physics} \textbf{1982},
  \emph{15}, 22 L717.

\bibitem{widom:physletta1982}
A.~Widom,
\newblock \emph{Physics Letters A} \textbf{1982}, \emph{90}, 9 474.

\bibitem{ceresoli:prb2006}
D.~Ceresoli, T.~Thonhauser, D.~Vanderbilt, R.~Resta,
\newblock \emph{Phys. Rev. B} \textbf{2006}, \emph{74} 024408.

\bibitem{xiao:rmp2010}
D.~Xiao, M.-C. Chang, Q.~Niu,
\newblock \emph{Rev. Mod. Phys.} \textbf{2010}, \emph{82} 1959.

\bibitem{solovyev:prb1997}
I.~V. Solovyev,
\newblock \emph{Phys. Rev. B} \textbf{1997}, \emph{55} 8060.

\bibitem{sivadas:prl2016}
N.~Sivadas, S.~Okamoto, D.~Xiao,
\newblock \emph{Phys. Rev. Lett.} \textbf{2016}, \emph{117} 267203.

\bibitem{noda:pccp2016}
Y.~Noda, K.~Ohno, S.~Nakamura,
\newblock \emph{Phys. Chem. Chem. Phys.} \textbf{2016}, \emph{18} 13294.

\bibitem{okugawa:jpcm2018}
T.~Okugawa, K.~Ohno, Y.~Noda, S.~Nakamura,
\newblock \emph{Journal of Physics: Condensed Matter} \textbf{2018}, \emph{30},
  7 075502.

\bibitem{smejkal:sciadv2020}
L.~Šmejkal, R.~González-Hernández, T.~Jungwirth, J.~Sinova,
\newblock \emph{Science Advances} \textbf{2020}, \emph{6}, 23 eaaz8809.

\bibitem{naka:natcomm2019}
M.~Naka, S.~Hayami, H.~Kusunose, Y.~Yanagi, Y.~Motome, H.~Seo,
\newblock \emph{Nature Communications} \textbf{2019}, \emph{10} 4305.

\bibitem{ahn:prb2019}
K.-H. Ahn, A.~Hariki, K.-W. Lee, J.~Kune\ifmmode~\check{s}\else \v{s}\fi{},
\newblock \emph{Phys. Rev. B} \textbf{2019}, \emph{99} 184432.

\bibitem{hayami:jpscp2020}
S.~Hayami, Y.~Yanagi, M.~Naka, H.~Seo, Y.~Motome, H.~Kusunose,
\newblock \emph{JPS Conf. Proc.} \textbf{2020}, \emph{30} 011149.

\bibitem{hayami:jpsj2019spinsplit}
S.~Hayami, Y.~Yanagi, H.~Kusunose,
\newblock \emph{Journal of the Physical Society of Japan} \textbf{2019},
  \emph{88}, 12 123702.

\bibitem{hayami:prb2020spinsplit}
S.~Hayami, Y.~Yanagi, H.~Kusunose,
\newblock \emph{Phys. Rev. B} \textbf{2020}, \emph{102} 144441.

\bibitem{yuan:prb2020}
L.-D. Yuan, Z.~Wang, J.-W. Luo, E.~I. Rashba, A.~Zunger,
\newblock \emph{Phys. Rev. B} \textbf{2020}, \emph{102} 014422.

\bibitem{naka:prb2021perovskite}
M.~Naka, Y.~Motome, H.~Seo,
\newblock \emph{Phys. Rev. B} \textbf{2021}, \emph{103} 125114.

\bibitem{yuan:prm2021}
L.-D. Yuan, Z.~Wang, J.-W. Luo, A.~Zunger,
\newblock \emph{Phys. Rev. Mater.} \textbf{2021}, \emph{5} 014409.

\bibitem{yuan:prb2021}
L.-D. Yuan, Z.~Wang, J.-W. Luo, A.~Zunger,
\newblock \emph{Phys. Rev. B} \textbf{2021}, \emph{103} 224410.

\bibitem{hernandez:prl2021}
R.~Gonz\'alez-Hern\'andez, L.~\ifmmode~\check{S}\else \v{S}\fi{}mejkal,
  K.~V\'yborn\'y, Y.~Yahagi, J.~Sinova, T.~c.~v. Jungwirth,
  J.~\ifmmode~\check{Z}\else \v{Z}\fi{}elezn\'y,
\newblock \emph{Phys. Rev. Lett.} \textbf{2021}, \emph{126} 127701.

\bibitem{smejkal:prx2022magnetoresistance}
L.~\ifmmode~\check{S}\else \v{S}\fi{}mejkal, A.~B. Hellenes,
  R.~Gonz\'alez-Hern\'andez, J.~Sinova, T.~Jungwirth,
\newblock \emph{Phys. Rev. X} \textbf{2022}, \emph{12} 011028.

\bibitem{mazin:pnas2021}
I.~I. Mazin, K.~Koepernik, M.~D. Johannes, R.~Gonz{\'a}lez-Hern{\'a}ndez,
  L.~{\v{S}}mejkal,
\newblock \emph{Proceedings of the National Academy of Sciences} \textbf{2021},
  \emph{118}, 42 e2108924118.

\bibitem{smejkal:prx2022spingroup1}
L.~\ifmmode~\check{S}\else \v{S}\fi{}mejkal, J.~Sinova, T.~Jungwirth,
\newblock \emph{Phys. Rev. X} \textbf{2022}, \emph{12} 031042.

\bibitem{smejkal:prx2022spingroup2}
L.~\ifmmode~\check{S}\else \v{S}\fi{}mejkal, J.~Sinova, T.~Jungwirth,
\newblock \emph{Phys. Rev. X} \textbf{2022}, \emph{12} 040501.

\bibitem{cheong:npj2025AMclassification}
S.-W. Cheong, F.-T. Huang,
\newblock \emph{npj Quantum Materials} \textbf{2025}, \emph{10}, 1 38.

\bibitem{guo:advmat2025review}
Z.~Guo, X.~Wang, W.~Wang, G.~Zhang, X.~Zhou, Z.~Cheng,
\newblock \emph{Advanced Materials} \textbf{2025}, 2505779.

\bibitem{hu:prx2025}
M.~Hu, X.~Cheng, Z.~Huang, J.~Liu,
\newblock \emph{Phys. Rev. X} \textbf{2025}, \emph{15} 021083.

\bibitem{provost:cmp1980}
J.~P. Provost, G.~Vallee,
\newblock \emph{Communications in Mathematical Physics} \textbf{1980},
  \emph{76}, 3 289 .

\bibitem{berry:book1989}
M.~V. Berry,
\newblock \emph{{The Quantum Phase, Five Years After}},
\newblock Advanced series in mathematical physics. World Scientific Publishing
  Company, \textbf{1989}.

\bibitem{resta:epjb2011}
R.~Resta,
\newblock \emph{The European Physical Journal B} \textbf{2011}, \emph{79}, 2
  121.

\bibitem{gao:prl2014BCP}
Y.~Gao, S.~A. Yang, Q.~Niu,
\newblock \emph{Phys. Rev. Lett.} \textbf{2014}, \emph{112} 166601.

\bibitem{SI}
{See Supporting Information.}

\bibitem{onishi:nanolett2025}
Y.~Onishi, H.~Isobe, A.~Shitade, N.~Nagaosa,
\newblock \emph{Nano Letters} \textbf{2025}, \emph{25}, 7 2763, pMID: 39927604.

\bibitem{lapa:prb2019}
M.~F. Lapa, T.~L. Hughes,
\newblock \emph{Phys. Rev. B} \textbf{2019}, \emph{99} 121111.

\bibitem{gao:prl2019}
Y.~Gao, D.~Xiao,
\newblock \emph{Phys. Rev. Lett.} \textbf{2019}, \emph{122} 227402.

\bibitem{marzari:prb1997}
N.~Marzari, D.~Vanderbilt,
\newblock \emph{Phys. Rev. B} \textbf{1997}, \emph{56} 12847.

\bibitem{marzari:rmp2012}
N.~Marzari, A.~A. Mostofi, J.~R. Yates, I.~Souza, D.~Vanderbilt,
\newblock \emph{Rev. Mod. Phys.} \textbf{2012}, \emph{84} 1419.

\bibitem{karplus:pr1954AHE}
R.~Karplus, J.~M. Luttinger,
\newblock \emph{Phys. Rev.} \textbf{1954}, \emph{95} 1154.

\bibitem{haldane:prl2004}
F.~D.~M. Haldane,
\newblock \emph{Phys. Rev. Lett.} \textbf{2004}, \emph{93} 206602.

\bibitem{nagaosa:rmp2010AHE}
N.~Nagaosa, J.~Sinova, S.~Onoda, A.~H. MacDonald, N.~P. Ong,
\newblock \emph{Rev. Mod. Phys.} \textbf{2010}, \emph{82} 1539.

\bibitem{gao:prb2015}
Y.~Gao, S.~A. Yang, Q.~Niu,
\newblock \emph{Phys. Rev. B} \textbf{2015}, \emph{91} 214405.

\bibitem{resta:arxiv2017DWOAM}
R.~Resta,
\newblock {Geometrical meaning of the Drude weight and its relationship to
  orbital magnetization}, \textbf{2017},
\newblock \urlprefix\url{https://arxiv.org/abs/1703.00712}.

\bibitem{resta:jpcm2018DWSCW}
R.~Resta,
\newblock \emph{Journal of Physics: Condensed Matter} \textbf{2018}, \emph{30},
  41 414001.

\bibitem{ashcroft}
N.~W. Ashcroft, N.~D. Mermin,
\newblock \emph{{Solid State Physics}},
\newblock Holt, Rinehart and Winston, New York, \textbf{1976}.

\bibitem{xiao:prl2005}
D.~Xiao, J.~Shi, Q.~Niu,
\newblock \emph{Phys. Rev. Lett.} \textbf{2005}, \emph{95} 137204.

\bibitem{xiao:prl2006thermoele}
D.~Xiao, Y.~Yao, Z.~Fang, Q.~Niu,
\newblock \emph{Phys. Rev. Lett.} \textbf{2006}, \emph{97} 026603.

\bibitem{thonhauser:prl2005}
T.~Thonhauser, D.~Ceresoli, D.~Vanderbilt, R.~Resta,
\newblock \emph{Phys. Rev. Lett.} \textbf{2005}, \emph{95} 137205.

\bibitem{shi:prl2007}
J.~Shi, G.~Vignale, D.~Xiao, Q.~Niu,
\newblock \emph{Phys. Rev. Lett.} \textbf{2007}, \emph{99} 197202.

\bibitem{thonhauser:ijmpb2011}
T.~THONHAUSER,
\newblock \emph{International Journal of Modern Physics B} \textbf{2011},
  \emph{25}, 11 1429.

\bibitem{souza:prb2000SWM_sumrule}
I.~Souza, T.~Wilkens, R.~M. Martin,
\newblock \emph{Phys. Rev. B} \textbf{2000}, \emph{62} 1666.

\bibitem{onishi:prx2024}
Y.~Onishi, L.~Fu,
\newblock \emph{Phys. Rev. X} \textbf{2024}, \emph{14} 011052.

\bibitem{tahir:prl2023}
M.~Tahir, H.~Chen,
\newblock \emph{Phys. Rev. Lett.} \textbf{2023}, \emph{131} 106701.

\bibitem{shitade:prb2025}
A.~Shitade,
\newblock \emph{Phys. Rev. B} \textbf{2025}, \emph{112} 174431.

\bibitem{sinova:prl2004}
J.~Sinova, D.~Culcer, Q.~Niu, N.~A. Sinitsyn, T.~Jungwirth, A.~H. MacDonald,
\newblock \emph{Phys. Rev. Lett.} \textbf{2004}, \emph{92} 126603.

\bibitem{mook:prr2020}
A.~Mook, R.~R. Neumann, A.~Johansson, J.~Henk, I.~Mertig,
\newblock \emph{Phys. Rev. Res.} \textbf{2020}, \emph{2} 023065.

\bibitem{roig:prb2024}
M.~Roig, A.~Kreisel, Y.~Yu, B.~M. Andersen, D.~F. Agterberg,
\newblock \emph{Phys. Rev. B} \textbf{2024}, \emph{110} 144412.

\bibitem{antonenko:prl2025}
D.~S. Antonenko, R.~M. Fernandes, J.~W.~F. Venderbos,
\newblock \emph{Phys. Rev. Lett.} \textbf{2025}, \emph{134} 096703.

\bibitem{xiao:prl2022hkspaceBCP}
C.~Xiao, H.~Liu, W.~Wu, H.~Wang, Q.~Niu, S.~A. Yang,
\newblock \emph{Phys. Rev. Lett.} \textbf{2022}, \emph{129} 086602.

\end{thebibliography}
\end{document}